\documentclass[referee]{raa}            

\usepackage{graphicx,times}             
\usepackage[fleqn]{amsmath}
\usepackage{color}
\usepackage{ulem}

\begin{document}

   \title{A strange phenomenon of XZ Andromedae: two Keplerian periods with a 1:3 ratio}

   \volnopage{Vol.0 (200x) No.0, 000--000}      
   \setcounter{page}{1}          

   \author{Jin-Zhao Yuan\inst{1} and Sheng-Bang Qian\inst{2}}
   \institute{Department of Physics, Shanxi Normal University, Linfen
041004, Shanxi, China; {\it yuanjz@sxnu.edu.cn}\\
     \and
     Yunnan Observatory, Chinese Academy of Sciences, PO Box 110, 650011 Kunming, China\\}

   \date{Received~~2018 month day; accepted~~2018~~month day}

\abstract{Six mid-eclipse times of the eclipsing binary XZ And are obtained, which are analysed together with others collected from the literature. Two sets of cyclic variations with periods of 33.43 and 100.4 yr are found if a double-Keplerian model is used to fit the data. The 1:3 ratio of the periods suggests that both cyclic variations arise from dynamic motions of two companions rather than magnetic activity of the eclipsing pair. According to the double-Keplerian model, the companions have the masses of $\sim1.32 M_{\bigodot}$ and $\sim1.33 M_{\bigodot}$, respectively. Comparing the total masses of the eclipsing pair of $3.12 M_{\bigodot}$, it is obvious that XZ And is a general N-body system. The strong gravitational perturbation between two companions invalidates the double-Keplerian model. It is strange that two Keplerian periods with a 1:3 ratio are derived from the best fits with the inappropriate model. The illogical, but interesting phenomena also appear in other two Algol systems, suggesting that our discoveries deserve attention from astronomers.
}

   \authorrunning{J.-Z. Yuan \& S.-B. Qian}            
   \titlerunning{Two periods with a 1:3 ratio}  
   \maketitle


%
%
\section{Introduction}           
\label{sect:intro}

XZ Andromedae (BD+$41^{\circ}367$) is a classic Algol-type binary (hereafter XZ And AB). The primary is a Main Sequence star and the more evolved secondary fills its Roche lobe (Manzoori 2016). Dugan and Wright (1939) found that primary eclipsing times and secondary eclipsing times follow a same linear ephemeris roughly, but they also pointed out quite intricate variations in the residuals (i.e., $O-C$) between observed ($O$) and computed ($C$) mid-eclipse times. Odinskaya and Ustinov (1952) found these irregular variations contain two sets of cyclic modulation. Todoran (1967) fitted the data between the years 1891-1919 with a sinusoidal curve with a period of 21.3 yr, and the data between 1924-1966 with another sinusoidal curve with a period of 44.6 yr. Todoran (1967) interpreted these variations as apsidal motion, but Kreiner (1976) disproved this explanation. Demircan et al. (1995) reported three cyclic variations with periods of 11.2 yr, 36.8 yr, and 137.5 yr, respectively, and attributed the cyclic variations to magnetic activity of the secondary and light-travel time (LTT) effect due to one or two under-luminous star(s) around the eclipsing pair. Frieboes-Conde \& Herczeg (1973) and Borkovits \& Heged\"{u}s (1996) claimed that an unambigous identification of third component was not possible for XZ And, but the latter still presented the orbital parameters of third and fourth bodies, including the periods of 35.6 yr and 69.8 yr, respectively. Selam and Demircan (1999) reported two periods of 36.79 and 126.35 yr for these two companions. Recently, Yang (2013) found only a quasi-cyclic period of 32.60 yr, while Manzoori (2016) obtained two periods, i.e., 23.3 and 34.8 yr. Just as before, both authors did not affirm the explicit reason for the variations. Therefore, it is necessary to reanalyze the behavior of the mid-eclipse times.

 All available mid-eclipse times collected from the literature and several new data obtained in this paper are used to plotted $O-C$ diagram in Section 2. In Section 3, the fitting procedures are described, and the best-fit solution is given. In Section 4, we summarize our results and give our conclusions in Section 5.

\section{Eclipse-timing variations}
\label{sect:ec-ti}

\renewcommand{\thefootnote}{\arabic{footnote}}
CCD photometric observations were carried out in the past six years. The 85-cm telescope at the Xinglong Station of National Astronomical Observatory of China (NAOC-85), equipped with a primary-focus multicolor CCD photometer (Zhou et al. 2009), was used in 2013 February and December. The 60-cm (YNAO-60) and 100-cm (YNAO-100) Cassegrain telescopes at Yunnan Observatory were used in 2013 January, 2015 February, 2016 February, and 2018 February. The comparison and check stars are GSC 02824-01778 ($\alpha_{J2000.0}$ = $01^{h}57^{m}14.^{s}2$, $\delta_{J2000.0}$ = $+42^{\circ}02^{\prime}19.^{\prime\prime}2$) and 2MASS
01564776+4201523 ($\alpha_{J2000.0}$ = $01^{h}56^{m}47.^{s}7$, $\delta_{J2000.0}$ = $+42^{\circ}01^{\prime}52.^{\prime\prime}8$), respectively. We used the aperture photometry package \textsc{iraf}{\footnote[1]{\textsc{iraf} is developed by the National
Optical Astronomy Observatories, which are operated by the Association of Universities for Research in Astronomy, Inc., under contract to the National Science Foundation.}} to reduce the CCD data. Six new mid-eclipse times are obtained by using a parabolic fitting method. The new data are listed in Table 1. A mean time is given if multi-band values were obtained simultaneously.

\begin{table*}
\begin{minipage}{12cm}
\caption{Six new mid-eclipse times of XZ And.}
\begin{tabular}{ccccc}\hline
HJD (UTC) & BJD (TDB) & Errors & Filters &  Origin\\
2400000+     &  2400000+   &  (d)  &     & \\\hline
56313.1564   & 56313.15716  & $\pm0.0002$ &       $R$ &   YNAO-100\\
56338.94475  & 56338.94552  & $\pm0.00005$ &     $V$ &  NAOC85\\
56655.19173   & 56655.19250  & $\pm0.00005$ &     $R$ &  NAOC85\\
57061.0204   & 57061.02119  & $\pm0.0002$ &      $R$ &  YNAO-100\\
57422.05866  & 57422.05947  & $\pm0.00009$ &     $V$ &  YNAO-100\\
57422.05866   & 57422.05947  & $\pm0.00011$ &     $R$ &  YNAO-100\\
58159.0693    & 58159.07013  & $\pm0.0001$ &      $V$ &  YNAO-60\\
58159.0693    & 58159.07013  & $\pm0.0001$ &      $R$ &  YNAO-60\\
\hline
\end{tabular}
\end{minipage}
\end{table*}

The Lichtenknecker Database of the BAV{\footnote[2]{http://www.bav-astro.de/index.php?sprache=en}} and the O-C Gateway Database{\footnote[3]{http://var.astro.cz/ocgate/}} list a large number of mid-eclipse times of XZ And, which come mainly from Zessewitsch (1924), Banachiewicz (1925), Dugan and Wright (1939), Kordylewska (1931), Lause (1934, 1936, 1949), Szafraniec (1950, 1952a, 1952b, 1955, 1957), Szczepanowska (1950, 1953, 1956, 1959), Piotrowski (1950), Odinskaya and Ustinov (1952), Ashbrook (1952a, 1952b ,1953), Domke \& Pohl (1953), Pohl (1955), Rudolph (1960), Robinson (1965a, 1965b, 1966, 1967a, 1976b), Todoran (1967, 1968, 1973), Todoran \& Popa (1967), Robinson \& Ashbrook (1968), Frieboes-Conde \& Herczeg (1973), Baldwin (1973, 1976, 1977, 1978), Mallama et al. (1977), Kreiner et al. (1980), Olson (1981), Baldwin \& Samolyk (1993), Heged\"{u}s et al. (1996), Agerer \& Huebscher (2003), Cook et al. (2005), H\"{u}bscher et al. (2005, 2006, 2009), Nagai (2007, 2008, 2010), Samolyk (2008, 2009, 2010), and Yang (2013).

Three visual times (HJD 2423681.21, 2423694.29, and 2423699.40) are discarded due to their poor precision, six mid-eclipse times (HJD 2423670.430, 2423756.292, 2441650.291, 2441958.429, 2444488.400, and 2450752.310) are not adopted for their large deviation from the $O-C$ curve. Finally, we have collected 1131 mid-eclipse times over a 127-year timespan. Most photographic and visual data were published without uncertainties, the typical uncertainty of $\sigma=\pm0.003$ d is used.  For CCD data, the uncertainty of $\pm0.0001$ d is adopted if it is less than $\pm0.0001$ d.

Usually, the mid-eclipse times were reported in the Heliocentric Julian Dates (HJD) based on Coordinated Universal Time (UTC)
standard, which is not strictly uniform. Therefore, we adopted the Barycentric Dynamical Time (TDB) standard, and corrected all data to Solar-system barycenter, giving Barycentric Julian Dates (BJD) (Eastman, Siverd \& Gaudi 2010). The relation between the Universal Time (UT) and the Terrestrial Time (TT) given by Duffett-Smith \& Zwart (2011) was used to convert the old data before 1950.

The calculated mid-eclipse epoches are computed with the linear ephemeris
\begin{equation}
\mathrm{BJD}2452500.51473 + 1^{d}.3572855 \times{E},
\end{equation}
where the period was also used by Manzoori (2016). In Equation (1), $E$ is the eclipse cycle number counted from $\mathrm{BJD}2452500.51473$.
We can calculated the residuals $O-C$, i.e., the observed mid-eclipse times minus the calculated mid-eclipse epoches. Figure 1 shows all $O-C$ values.

\begin{figure}
\vspace{0.5cm}
\begin{center}
\includegraphics[width=8.5cm]{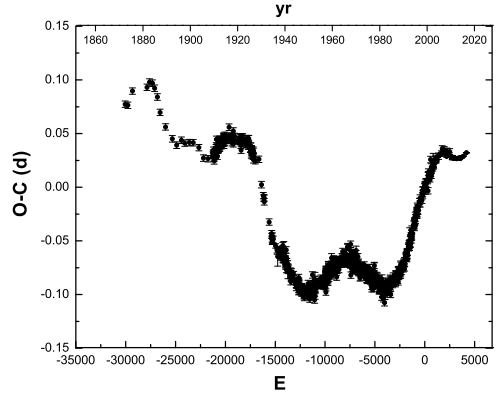}
\caption{The $O-C$ diagram of XZ And.}\label{fig1}
\end{center}
\end{figure}

\section{Data analysis and LTT models}

The secondary component is transfering mass to the primary. Therefore, the observed period increases (Yang 2013; Manzoori 2016), and the $O-C$ curve should have a parabolic trend. Figure 1 shows that an additional periodic model is also required.
Following the method adopted by Yuan et al. (2016), we first use a quadratic plus one-companion model
\begin{equation}
O-C = T_{O}(E) - T_{C}(E) = C_0 + C_1\times{E}+ C_2\times{E^2} + {\tau}_3
\end{equation}
to fit the $O-C$ values. The LTT term, ${\tau}_3$, arises from the variation of distance of an eclipsing binary from the observer as a result of a distant third component, can be calculated using the following equation (Irwin 1952)
\begin{equation}
{\tau}_3=\frac{a_{3}\sin
i_3}{c}\Big[\frac{1-{e_3}^2}{1+e_3\cos\nu_3}\sin(\nu_3+\omega_3)+e_3\sin\omega_3\Big],
\end{equation}
where $a_{3}{\sin}i_{3}$ is the semi-major axis of the eclipsing binary around the barycentre of the triple system, projected onto the tangent plane of the sky. $\omega_3$ is the argument of the periastron measured from the ascending node and $e_3$ the eccentricity. For any mid-eclipse time $t$, the true anomaly $\nu_3$, can be derived from the following relation
\begin{equation}
\mathrm{tan}~\frac{{\nu}_3}{2}  =
\sqrt{\frac{1+e_3}{1-e_3}}~\mathrm{tan}~\frac{\varphi_3}{2}
\end{equation}
where $\varphi_3$ is the eccentric anomaly, and can be obtained by solving the Kepler's equation
\begin{equation}
M_3 = \varphi_3 - e_3~\mathrm{sin}~\varphi_3,
\end{equation}
In Equation (5), the mean anomaly $M_3=2\pi(t-T_3)/P_3$, where $T_3$ is the time of the periastron
passage, and $P_3$ is the orbital period.

For one ($e_3$, $T_3$, $P_3$) configuration, we fit the $O-C$ data with equation (2), and get the goodness-of-fit statistic
\begin{equation}
\chi^2 = \sum_{i=1}^{1131}\Big[\frac{y_i-y(t_i)}{{\sigma}_i}\Big]^2,
\end{equation}
where $y_i$ is the $O-C$ value given by Equation (1), and $y(t_i)$ is the model value at mid-eclipse time $t_i$ calculated by Equation (2).
In equation (6), $\sigma_i$ is the uncertainty of the $O-C$ data $y_i$ (i.e., the uncertainty of the $i$-th mid-eclipse time). The best $a_{3}{\sin}i_{3}$ and $\omega_3$ can be obtained from the best fit. Searching $e_3$ from 0.0 to 0.99, and $T_3$ from 24500000.0 to 24500000.0 + $P_3$, the local $\chi^2$ minimum is obtained for the particular $P_3$, i.e., $\chi^{2}(P_3)$, which is plotted in Figure 2.
\begin{figure}
\vspace{0.5cm}
\begin{center}
\includegraphics[width=8.5cm]{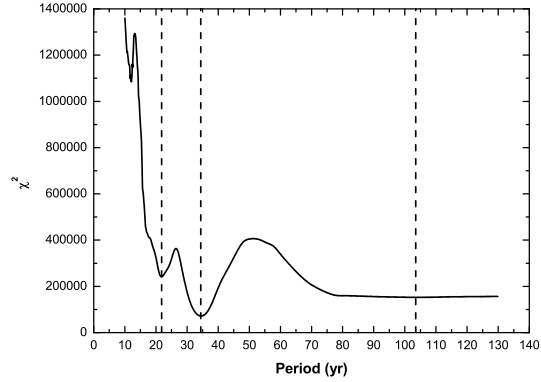}
\caption{The Keplerian periodogram of XZ And. The dashed vertical lines mark three $\chi^2$ minima.}\label{fig2}
\end{center}
\end{figure}

Figure 2 shows that $\chi^2$ reaches the minimum at $P = 34.48$ yr, suggesting a companion with a period of 34.48 yr. Hereafter, we refer to the eclipsing pair as XZ And AB, and the companion as XZ And (AB)C. The best fits corresponding to the 34.48 yr periodicity are plotted in Figure 3, and listed in the second column (Solution 1) of Table 2.  As shown in the bottom panel of Figure 3, most data show residuals larger than $\pm$0.01 d, which is much larger than the typical uncertainty, i.e., $\pm$0.003 d.

\begin{figure}
\vspace{0.5cm}
\begin{center}
\includegraphics[width=8.5cm]{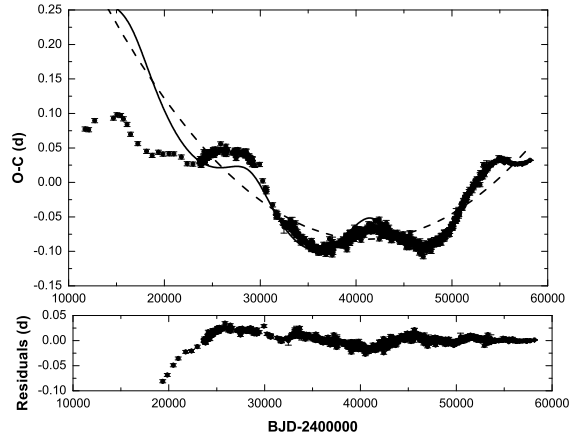}
\caption{The best fit to the eclipse-timing variations of XZ And with the one-companion model. The overplotted solid line denotes the best fit with Equation (2), and the dashed line only represents the second-order polynomial in the ephemeris. The residuals of the best fit are displayed in the lower panel. Note that the residuals before BJD2420000 reach as large as -0.22 d, and not appear in the diagram.}\label{fig3}
\end{center}
\end{figure}

In Figure 2, the one-companion fit shows another periodicity at $>$75 yr, suggesting another companion (XZ And (AB)D) with a longer period. But, due to the short time coverage and low precision, $\chi^2$ remains at very low level beyond 75 yr. We use a parabola plus two-companion model to fit the $O-C$ data. Just as the best fit with the one-companion model, we fix $e_3$, $T_3$, $P_3$, $e_4$, $T_4$, and $P_4$ during the fitting process. The parameters with the subscript `$4$' are similar to those with the subscript `$3$', but refer to the barycentre of XZ And AB and C around the barycentre of XZ And AB, C and D. After searching all possible $e_{3,4}$ and $T_{3,4}$, we obtain the local $\chi^2$ minimum for the fixed $P_3$ and $P_4$, i.e. $\chi^{2}(P_3, P_4)$. $\chi^{2}(P_3, P_4)$ is a function of $P_3$ and $P_4$. We search $P_3$ in 20-40 yr, and $P_4$ in 70-120 yr simultaneously. Finally, a two-dimensional periodogram results, and is shown in Figure 4. The global $\chi^2$ minimum is located at ($P_4\simeq97.8$ yr, $P_3\simeq33.4$ yr), which confirms XZ And (AB)C and D.

\begin{figure}
\vspace{0.5cm}
\begin{center}
\includegraphics[width=8.5cm]{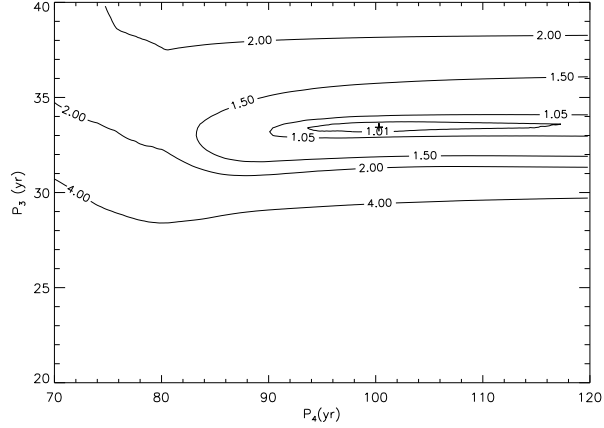}
\caption{Two-dimensional periodogram of XZ And derived from a parabola plus two-companion model. The $\chi^2$ contours have been normalized by division of the global ${\chi}^{2}$ minimum, which is marked by a cross.}\label{fig4}
\end{center}
\end{figure}

Since we search $P_3$ and $P_4$ in finite step (i.e., 0.2 day), the global $\chi^2$ minimum derived from the two-dimensional periodogram is not the true minimum, but very close to the true minimum. Starting from the "best" solution in the two-dimensional periodogram, we fit the data by using Levenberg-Marquardt fitting algorithm (Markwardt 2009). The Levenberg-Marquardt fits set all parameters free. The free parameters are $C_0$, $C_1$, $C_2$, $P_k$, $T_k$, $e_k$, $A_k$, $B_k$ ($k$ = 3, 4), where $A_k$ and $B_k$ are related to $a_k\sin i_k$ and $\omega_k$ (see Yuan \& \c{S}enavc{\i} 2014 for details). The best parameters and the least $\chi^2$ are listed in the fourth column (i.e., Solution 3) of Table 2. Interestingly, $P_4$ and $P_3$ are 100.3$\pm1.5$ yr and 33.43$\pm0.03$ yr, repsectively, suggesting a possible mean-motion resonance. The improved fits are plotted in Figure 5. Just as shown by Figure 5, most of the residuals are within $\pm$0.01, and much better than that of Solution 1. The best fit fails before BJD2420000, and around BJD2432000, where the data are scarce. We remind the reader that the old visual data before A.D. 1900 are much low-precious, and can not be fitted very well in most cases, such as SW Lac (Yuan \& \c{S}enavc{\i} 2014) and Z Dra (Yuan et al 2016). Although the ${\chi}^{2}$ statistic is relatively poor (the reduced chi-square statistic ${\chi}_{\nu}^{2}=15.2$), there is a good qualitative correspondence between the morphologies of the observed and model curve. In Figure 1., the thick $O-C$ curve shows that most visual data often conflict with each other within their typical uncertainties, i.e., $\pm$0.003, and only seem consistent within $\pm$0.01. Perhaps this explains why ${\chi}_{\nu}^{2}$ is large.

The uncertainty of $\sigma=\pm0.004$ d is also used for the photographic and visual data which were published without uncertainties. We refit the $O-C$ data, and obtain similar results (see Solution 4 in Table 2). XZ And (AB)D has an orbital periods of $P_5=102.9\pm2.4$ yr, while $P_4=33.34\pm0.03$ yr for XZ And (AB)C.

Figure 2 indicates that a period of $\sim$ 23 yr is also possible. It is likely that such LTT signal also appears in the bottom panel of Figure 5. It seems that a short-period companion (XZ And (AB)E) exists. For safety, the two-companion model is used again. This time, $P_4$ is still searched around 33 yr, but $P_3$ around 23 yr. To avoid confusion, the subscript '5' is used for XZ And (AB)E, while '3' for XZ And (AB)C. The Levenberg-Marquardt fit gives Solution 2, which is listed in the third column of Table 2. XZ And (AB)E has an orbital period of $P_5=\sim 24.35$ yr and a mass of $0.34M_{\bigodot}$. XZ And (AB)E produces a cyclic $O-C$ variation with a semi-amplitude of $a_5\sin i_5=1.25$ au, which is much smaller than $a_3\sin i_{3}$ and $a_4\sin i_{4}$. Compared to Solution 2, the $\chi^2$ in Solution 3 is much smaller, suggesting that Solution 3 is better. In Figure 4, the two-dimensional periodogram also reveals that the configure of ($P_4\simeq100$ yr, $P_3\simeq33$ yr) is more likely than that of ($P_4\simeq100$ yr, $P_5\simeq24$ yr). Therefore, we infer that such small signal may arise from unavoidable and slight imperfection in the double-Keplerian model.

\begin{figure}
\vspace{0.5cm}
\begin{center}
\includegraphics[width=8.5cm]{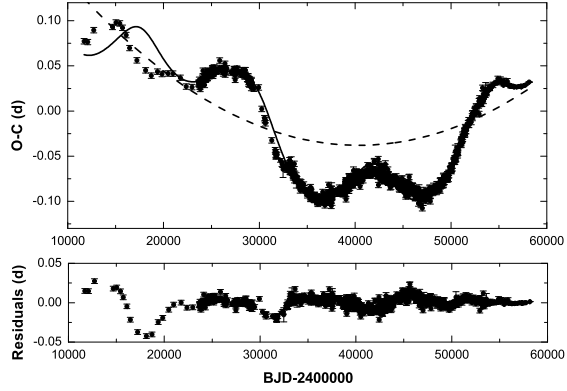}
\caption{The two-companion fit to the eclipse-timing variations of XZ And. The residuals of the best fit are displayed in the lower panel. The overplotted solid line denotes the best fit with a parabola plus two-companion model, and the dashed line only represents the parabola.}\label{fig5}
\end{center}
\end{figure}

\begin{table*}
\caption{The best-fit parameters of the companions around XZ And.}
\begin{tabular}{lcccc}\hline
parameter & Solution 1 &  Solution 2  &  Solution 3  &  Solution 4\\\hline
$C_0$ (d) & -0.0222$\pm0.0001$ & -0.0220$\pm0.0010$  & -0.0088$\pm0.0013$ & -0.0046$\pm0.0006$\\
$C_1$ ($\times10^{-5}$ d) & 1.43$\pm0.00$  & 1.47$\pm0.03$   & 0.65$\pm0.01$ & 0.70$\pm0.02$\\
$C_2$ ($\times10^{-10}$ d) & 8.48$\pm0.01$  &  8.63$\pm0.01$  & 3.65$\pm0.03$ & 3.75$\pm0.02$ \\\hline
$P_5$ (yr)  &    &  24.35$\pm0.03$     &   &\\
$T_5$ (BJD) &    &  2402255.0$\pm74.2$ &   &\\
$e_5$       &    &  0.638$\pm0.006$    &   &\\
$a_5\sin i_5$ (au)&    & 1.25$\pm0.02$  &   &\\
$\omega_5 (deg)$  &    & 149.2$\pm1.2$  &   &\\
$m_{5}$ ($M_{\bigodot}$, $i_5 = 89.8^{\circ}$) &   &  0.34$\pm0.01$ &  & \\
$A_{5}$ (au, $i_5 = 89.8^{\circ}$) &   &  12.71$\pm0.01$  &  & \\\hline
$P_3$ (yr) & 34.48$\pm0.01$  &  34.48$\pm0.03$  &  33.43$\pm0.03$ &  33.34$\pm0.03$\\
$T_3$ (BJD) & 2403941.0$\pm30.4$  &  2402431.6$\pm46.2$  & 2405738.2$\pm48.7$ & 2405981.9$\pm52.4$\\
$e_3$ &  0.256$\pm0.002$  &  0.174$\pm0.004$   &   0.228$\pm0.002$ &   0.221$\pm0.002$ \\
$a_3\sin i_3$ (au) & 5.30$\pm0.00$  &  4.45$\pm0.04$   & 5.09$\pm0.03$ & 5.15$\pm0.04$ \\
$\omega_3 (deg)$ &102.8$\pm0.8$  & 62.5$\pm1.4$   &   114.4$\pm1.1$ &   118.4$\pm1.2$\\
$m_{3}$ ($M_{\bigodot}$, $i_3 = 89.8^{\circ}$) &  1.36$\pm0.00$  &  1.16$\pm0.02$  & 1.33$\pm0.01$  &  1.35$\pm0.01$\\
$A_{3}$ (au, $i_3 = 89.8^{\circ}$) &  17.46$0.01$  &  17.65$\pm0.02$  &  17.06$\pm0.02$ &  17.06$\pm0.03$ \\\hline
$P_4$ (yr)  &    &      &  100.3$\pm1.5$  &  102.9$\pm2.4$ \\
$T_4$ (BJD) &    &   &  2426568.8$\pm43.9$ &  2426505.7$\pm51.5$\\
$e_4$       &    &     &  0.49$\pm0.01$ &  0.49$\pm0.01$ \\
$a_4\sin i_4$ (au)&    &   & 8.84$\pm0.10$ & 9.09$\pm0.16$ \\
$\omega_4 (deg)$  &    &   & 115.4$\pm1.7$ & 114.8$\pm2.8$ \\
$m_{4}$ ($M_{\bigodot}$, $i_4 = 89.8^{\circ}$) &  &  &1.32$\pm0.01$  & 1.34$\pm0.01$\\
$A_{4}$ (au, $i_4 = 89.8^{\circ}$) &  &  &38.70$\pm0.07$ &39.46$\pm0.11$\\\hline
$\chi^2$ & 71646.1  &  58605.5  & 16996.5 & 14758.4 \\\hline
\end{tabular}
\end{table*}

\section{RESULTS AND DISCUSSIONS}
\label{sect:result}
In this paper, new CCD observations of the Algol-type binary XZ And and all available mid-eclipse times in the literature are investigated.
The results are listed as Solution (3) in Table 2. The $O-C$ diagram shows a quadratic trend, suggesting that the orbital period of the eclispe binary increases with a rate of $dP/dt = 1.96 \times 10^{-7} \mathrm{d~yr^{-1}}$. By coincidence, Z Dra has a similar orbital period and increasing trend (Yuan et al. 2016). The increasing trend is attributed to mass transfer from the secondary component to the primary one. The mass transfer rate can be derived from the following equation
\begin{equation}
\dot{m_1}=\frac{m_{2}q}{3(1-q)}\frac{\dot{P}}{P}.
\end{equation}
For XZ And, Manzoori (2016) reported that $m_{1} = 2.10~\mathrm{M_{\odot}}$, $m_{2} = 1.02~\mathrm{M_{\odot}}$, and the mass ratio of the eclipsing pair $q = m_{2}/m_{1}=0.485$, giving the mass transfer rate of $\mathrm{d}m_{1}/\mathrm{d}t = 4.6\times10^{-8} \mathrm{M_{\bigodot}~yr^{-1}}$. The mass transfer rate is larger than that of Z Dra ($\mathrm{d}m_{1}/\mathrm{d}t = 9.2\times10^{-9} \mathrm{M_{\bigodot}~yr^{-1}}$), but often lower than those of contact binaries. Z Dra is an Algol-type binary with similar period as XZ And (Yuan et al. 2016). For contact binaries, such as AD Cnc (Qian et al. 2007a), V382 Cyg (Qian et al. 2007b), and TU Mus (Qian et al. 2007b), the typical value is $\sim10^{-7} \mathrm{M_{\bigodot}~yr^{-1}}$.

We find that the $O-C$ curve shows two sets of cyclic variations with periods of $33.43$ and $100.3$ yr, respectively. Interestingly, the ratio of the two periods is 1:3, or close to 1:3, which is a dynamical character. Although magnetic activity can explain biperiodic variations in the mid-eclipse times of an eclipsing binary (Applegate 1992; Yuan \& Qian 2007), magnetic activity can not produce two sets of variations with commensurate periods, especially for two periods with a (near) 1:3 ratio. The only reason for such variations is the light-travel time effect induced by two companions in a possible mean-motion resonance.

Manzoori (2016) carried out the photometric-spectroscopic analysis, and indicated that the orbital inclination of the eclipsing pair is $89.8^{\circ}$, and the total masses of the eclipsing pair are $m_{b}=3.12~M_{\bigodot}$. Assuming that the orbits of two companions are coplanar with the eclipsing pair, the minimum masses of two companions can be derived from the following mass functions
\begin{align}
& \frac{(m_3{\sin}i_3)^3}{(m_{b} + m_3)^2} = \frac{4\pi^2}{G{P_3}^2}\times(a_3\sin
i_3)^3,\\
& \frac{(m_4{\sin}i_4)^3}{(m_{b} + m_3 + m_4)^2} = \frac{4\pi^2}{G{P_4}^2}\times(a_4\sin
i_4)^3,
\end{align}
where the subscripts '3' and '4' refer to  XZ And (AB)C and D, respectively. The results reveals that XZ And (AB)C has the mass of $\sim1.33~M_{\bigodot}$, and the outer companion XZ And (AB)D $\sim1.32~M_{\bigodot}$. The semimajor axes of the orbits of XZ And (AB)C and D are $A_{3}=a_{3}\cdot(m_b+m_3)/m_3 = 17.06 ~au$ and $A_{4}=a_{4} \cdot (m_b+m_3+m_4)/m_4 = 38.70~au$, respectively. Obviously, XZ And is a general three-body system if the central eclipsing binary is treated as a single object. According to the double-Keplerian model, we can calculate the gravitational perturbation between two companions. For the inner companion, XZ And (AB)C, the ratio of the gravitational perturbation from the out companion to the centripetal forces from the eclipsing pair is between 0.03 and 0.19 with the average value of 0.085. For the outer companion, XZ And (AB)D, the average ratio of the gravitational perturbation from the inner companion to the centripetal forces from the eclipsing pair is 0.53. The strong gravitational perturbation invalidates the double-Keplerian model. However, it is strange that two interesting Keplerian periods are derived by using the inappropriate model.

The illogical, but interesting phenomena also appear in other two Algol systems. They are Z Dra (Yuan et al. 2016) and SW Lac (Yuan \& \c{S}enavc{\i} 2014). Yuan \& \c{S}enavc{\i} (2014) found that two companions are in a near 1:3 MMR orbits around the eclipsing binary SW Lac with periods of 82.6 and 27.0 yr. If the orbital inclinations of two companions of SW Lac are $90.0^{\circ}$, we can calculate the minimum masses of both companions ($m_3 = 0.62~M_{\bigodot}$, $m_4 = 1.90~M_{\bigodot}$) and the semimajor axes ($A_3 = 12.6~au$, $A_4 = 31.6~au$) from the best-fitting parameters (see Table 2 in Yuan \& \c{S}enavc{\i} 2014), whereas the total masses of the eclipsing pair are $m_{b}=2.13~M_{\bigodot}$. The gravitational perturbation between two companions is a little stronger than that of XZ And. Yuan et al. (2016) claimed that the Algol-type binary Z Dra has two companions with the periods of 59.88 and 29.96 yr, close to a 1:2 MMR. For Z Dra, $m_3 = 0.33~M_{\bigodot}$, $m_4 = 0.77~M_{\bigodot}$, $A_3 = 12.3~au$, $A_4 = 21.9~au$, and $m_{b}=1.90~M_{\bigodot}$ (see Table 2 in Yuan et al. 2016). Figure 7 presented by Yuan et al. (2016) shows that the gravitational perturbation is weaker than that of XZ And, but not yet ignorable.

The interesting phenomena can not appear in three Algol systems by chance. We infer that the interesting periods and the "inappropriate" double-Keplerian model reveal some unknown results. The results may be related with the dynamical characters of general N-body systems, or quantization of gravitation.

\normalem
\begin{acknowledgements}
This research has also made use of the Lichtenknecker-Database of the BAV, operated by the Bundesdeutsche Arbeitsgemeinschaft f\"{u}r Ver\"{a}nderliche Sterne e.V. (BAV). The computations were carried out at National Supercomputer Center in Tianjin, and the calculations were performed on TianHe-1(A). This work is supported by the Natural Science Foundation of China (NSFC) (11705111 and U1231121).
\end{acknowledgements}

\label{lastpage}

\end{document}